\documentclass[twocolumn,prl,aps,superscriptaddress,showpacs]{revtex4-1}

\usepackage{mathptmx}
\usepackage{psfrag,graphicx}
\usepackage{dcolumn}
\usepackage{amsmath,amssymb}
\usepackage{bm}
\usepackage{color}
\usepackage{latexsym}
\usepackage{color}
\usepackage[english]{babel}
\usepackage{latexsym}

\usepackage{amsmath} 
\usepackage{amssymb} 
\usepackage{amsfonts}
\usepackage{bm}
\usepackage{natbib}
\usepackage{epstopdf}
\usepackage{appendix}

\definecolor{mygrey}{gray}{0.35}
\definecolor{myblue}{rgb}{0.2,0.2,0.8}
\definecolor{myzard}{cmyk}{0,0,0.05,0}
\definecolor{mywhite}{rgb}{1,1,1}
\definecolor{mywhite}{rgb}{1,1,1}
\definecolor{myred}{rgb}{1,0.,0.3}

\usepackage[colorlinks=true,citecolor=myblue,linkcolor=myred]{hyperref}

\def\be{\begin{equation}}
\def\ee{\end{equation}}
\def\ba{\begin{align}}
\def\enda{\end{align}}
\def\bi{\begin{itemize}}
\def\ei{\end{itemize}}

 \def\ee{\mathord{\rm e}}

 \def\ee{\mathord{\rm e}}

\renewcommand{\ee}{{\rm e}}

\def\beq{\begin{equation}}
\def\beq{\begin{equation}}
\def\eeq{\end{equation}}

 \newcommand{\ket}[1]{|#1\rangle}

\def\II{{\rm I}}
\def\LL{{\rm L}}
\def\cc{{\rm c}}

\def\TTr{{\rm Tr}}

\newcommand{\ud}[1]{{#1^{\dagger}}}

\newcommand{\mean}[1]{\langle #1\rangle}

\begin{document}


\title{Mesoscopic Entanglement Induced by Spontaneous Emission in Solid-State Quantum Optics}

 \author{Alejandro Gonz\'alez-Tudela} 
 \affiliation{
 Departamento de F\'isica Te\'orica de la Materia Condensada,
 Universidad Autonoma de Madrid, 
 28040 Madrid, 
 Spain}

 \author{Diego Porras}
 \affiliation{
 Departamento de F\'isica Te\'orica I,
 Universidad Complutense, 
 28040 Madrid, 
 Spain}

\date{\today}

\begin{abstract}
Implementations of solid state quantum optics provide us with devices where qubits are placed at fixed positions in photonic or plasmonic one dimensional waveguides.
We show that solely by controlling the position of the qubits and with the help of a coherent driving, collective spontaneous decay may be engineered to yield an entangled mesoscopic steady-state. 
Our scheme relies on the realization of pure superradiant Dicke models by a destructive interference that cancels dipole-dipole interactions in one-dimension. 
\end{abstract}

\maketitle

The study of atoms coupled to the electromagnetic (e.m) field confined in cavities or optical waveguides has played a central role in the fields of Quantum Optics and Atomic Physics. 
In recent years the basics of that physical system has been realized with artificially designed atoms in solid-state setups. We may include here quantum dots and nitrogen vacancy (NV) centers deterministically coupled to photonic cavities  \cite{Badolato05sci,Imamoglu07nat,englund10a}, and  plasmonic \cite{Akimov07nat,huck11a} or photonic \cite{lonvcar00a,vlasov05a,schwoob05a,lundhansen08a,laucht12a} waveguides, as well as circuit QED setups where superconducting qubits are coupled to microwave cavities \cite{Wallraff04nat,Astafiev10sci}. Even though the physics of atomic and solid-state quantum optical systems is similar, the latter shows a crucial advantage: on a solid substrate, emitters may be placed permanently at fixed positions at separations of the order of relevant wavelengths \cite{mohan10a}. An important application of those systems is quantum information processing in solid devices \cite{Imamoglu99prl}, where artificial atoms acting as qubits are placed within the e.m. field confined in a 
microcavity. Typically, the realization of those ideas requires unitary qubit-
field evolutions and a natural candidate is to consider collective couplings to a single mode in a cavity. An alternative pathway is to tailor the interaction with the environment to induce quantum correlations between qubits with dissipation \cite{plenio99a,verstraete09a}.
This approach has been proven to be advantageous to generate entanglement between ensembles of atoms \cite{krauter11a}. 
In this direction one-dimensional guided modes have been recently pointed out as an useful tool to create two-qubit entanglement \cite{Gonzalez11prl,martincano11a,dzsotjan10a} and many-qubit entanglement in cascaded quantum networks \cite{Stannigel12a}. 

\begin{figure}[tb]
\begin{center}
\includegraphics[width=0.45\textwidth]{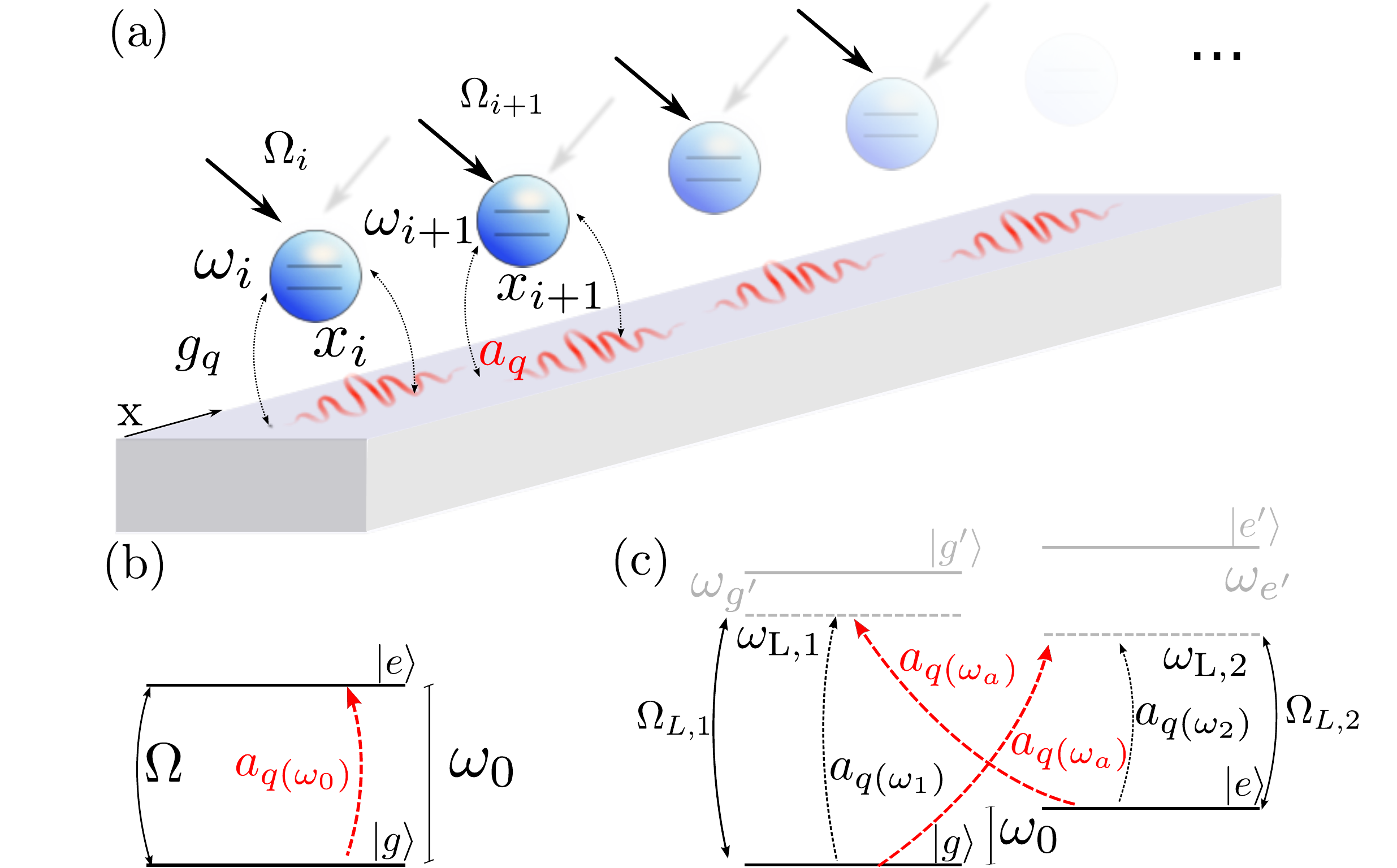}
\end{center}
\caption{(Color Online) (a): Experimental scheme of the system: ensemble of equally-spaced qubits placed in the vicinity of a one-dimensional waveguide. (b): Two-level system configuration with resonant excitation. Panel (c): Four-level system configuration with two-additional lasers, where we impose the condition: $\omega_{\LL,1} - \omega_0 = \omega_{\LL,2} + \omega_0=\omega_a$ and define $\omega_1=\omega_{\LL,1}$, $\omega_2=\omega_{\LL,2}-\omega_0$.}
\label{fig1}
\end{figure}

In this Letter, we show that by placing a set of qubits in a one-dimensional waveguide (see Fig. \ref{fig1}) the continuum of e.m. field modes induces a controllable dissipative coupling between the qubits. 
The possibility of deterministically position the artificial atoms or qubits allows us to engineer the paradigm for quantum optical collective effects, i.e. the Dicke model of superradiance \cite{Dicke54pr} in its pure form.
The observation of the latter in optical systems is hindered due to dephasing caused by dipole-dipole interactions \cite{Haroche82physrep,Lehmberg70pra}. In our scheme those interactions can be switched off by an appropriate choice of the inter-qubit distance. Adding a classical drive to the pure Dicke model we obtain a dissipative system with a phase diagram of steady states showing mesoscopic spin squeezing and entanglement. This model has been theoretically investigated in the past \cite{Drummond78oc,Drummond80pra, Milburn02pra}, but experimental realizations are scarce.
Finally we upgrade our scheme to a set of $N$ 4-level emitters \cite{Elzerman11a,Weiss12prl} and show that a judicious choice of couplings to the waveguide and dispersion relations may lead to a variety of many-body dissipative models which show entangled steady-states. 
%

We start by modeling $N$ 2-level systems (2LS), $\{ |g\rangle_n$, $| e \rangle_n \}_{n = 1 \dots N}$, placed at positions $x_n$ and coupled to a one dimensional field (Fig. \ref{fig1} (a)) with photon annihilation operators $a_q$, 
described by the Hamiltonian $H = H_0 + H_\II$. The free term is 
$H_0 = H_{\rm qb} + H_{\rm field}$, with ($\hbar = 1$)
\begin{equation}
H_{\rm qb} = \frac{\omega_0}{2}  \sum_{n=1}^N \sigma_n^{z}, \ \ \ \
H_{\rm field} = \sum_q \omega_q a^\dagger_q a_q\,,
\label{H0}
\end{equation}
where $\omega_0$ is the qubit energy (Fig \ref{fig1} (b)) 
and $\omega_q$ is the field dispersion relation. 
We define Pauli matrices $\sigma^z_n = |e \rangle_n \langle e| - |g \rangle_n \langle g|$, 
$\sigma^+_n = |e\rangle_n \langle g|$, $\sigma^-_n = |g\rangle_n \langle e|$.
The photon polarization is neglected to focus on the most relevant physics of our work.
We consider a dipolar coupling of the form
\begin{equation}
H_\II  = \sum_{n} \left( \sigma_n E(x_n) + \rm {H.c.} \right)\,,
\label{Hint}
\end{equation}
with $E(x) = \sum_q g_q (a_q e^{i q x}+\ud{a}_q e^{-i q x})$, and $g_q$ is a dipolar qubit-field coupling. 
Define $\rho$, the reduced density matrix for the qubits. 
In the weak coupling limit, the evolution of $\rho$ 
can be described by a markovian master equation of the form 
$d \rho /d t = {\cal L}(\rho)$ \cite{breuerbook}, with the superoperator
\begin{equation}
\label{mequation1}
{\cal L}(\rho) = 
\sum_{n,m} J_{n,m} \left( \sigma^-_n \rho \sigma^+_m - \rho \sigma^+_m \sigma^-_n \right)
+ \rm {H.c.}\, .
\end{equation}
A detailed derivation follows the description in dimensions higher than one presented in previous works \cite{Lehmberg70pra} (see Section A Sup. Mat.). Special care must be paid to the counter rotating terms in (\ref{mequation1}), which have to be included to get the following result for the collective decay rates
\begin{eqnarray}
J_{n,m} &=& \frac{\Gamma}{2} e^{i q(\omega_0) |x_n - x_m|}\,.
\label{col.rates}
\end{eqnarray}
We define $\Gamma = \gamma(\omega_0)$, with the function 
$\gamma(\omega) = g_{q(\omega)}^2 D(\omega)/\pi$, where 
$q(\omega)$ is the resonant  wavevector at $\omega$, and we have defined the e.m. density of states, $D(\omega) = (2 \pi / L) |d q(\omega)/d \omega|$, with $L$ the quantization length.
A crucial observation for this work, 
is that couplings $J_{n,m}$ ideally do not decay with the distance,
a situation that is singular of one dimensional waveguides. In free space on the contrary collective couplings decay like $1/r$ or $1/r^3$, depending on the relative dipole orientation \cite{Lehmberg70pra}.

Homogeneous couplings \cite{Gonzalez11prl,martincano10a,martincano11a,kien05a,chang12a} $J_{n,m} = \Gamma/2$ can be obtained from Eq. 
(\ref{col.rates}) by the choice $x_n = n \lambda_0$, with $\lambda_0 = 2 \pi / q_0$, and 
$n \in \mathbb{Z}$. This condition cancels dipole-dipole interactions and we get the pure Dicke superradiant decay described by 
\begin{equation}
{\cal L}_{\rm D}(\rho) = 
\frac{\Gamma}{2} \left(S^- \rho S^+ -  S^+ S^- \rho \right) + \rm H.c.\,,
\label{Dicke}
\end{equation}
with $S^- = \sum_n \sigma^-_n$, $S^+ = \sum_n \sigma^+_n$. We also define 
$S_\alpha = \sum_n \sigma^\alpha_n/2$ ($\alpha = x,y,z$), and the basis 
$\{| J, M \rangle \}$ of eigenstates of $\vec{S}^2$, $S_z$.
Assuming an initial state like 
$| \Psi_0 \rangle = \otimes_n | e \rangle_n = | N/2, N/2 \rangle$ the system evolves within the sector $J = N/2$. We note that Dicke superradiant decay is achieved in one dimension without the restriction that the whole qubit ensemble is confined within a region of length $\lambda_0$, which is a requirement for other realizations, i.e., atomic ensembles.

\begin{figure}[h]
\begin{center}
\includegraphics[width=0.4\textwidth]{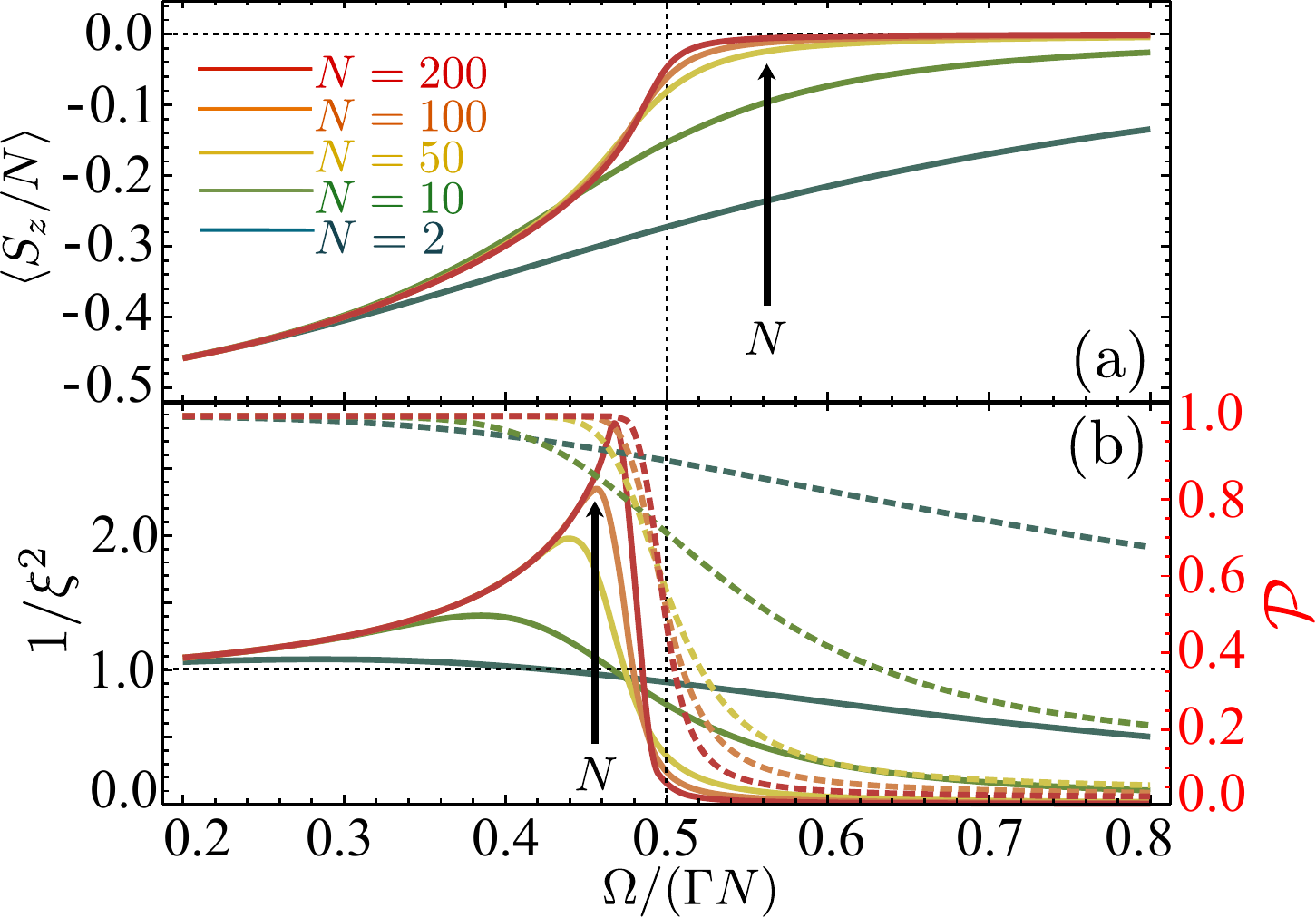}
\end{center}
\caption{(Color online) 
Numerical results for the coherently pumped Dicke model (\ref{Dicke.pumped}) for increasing $N$. 
Different colors represent different number of qubits from $N=2$ (blue) to $N=200$ (red) as shown in the legend.
(a) Population inversion $\mean{S_z}/N$. (b) Purity $\mathcal{P}$ (dashed) and 
spin-squeezing parameter $1/\xi^2$(solid).} 
\label{fig2}
\end{figure}

In this work we focus on the qubit steady-state, $\rho_{\rm s}$, which for a given Liouvillian fulfills ${\cal L}(\rho_{\rm s}) = 0$. To achieve some controllability on 
$\rho_{\rm s}$, we add a pump term which physically can be implemented by the interaction of qubits with a resonant field with Rabi frequency $\Omega$,
\begin{equation} 
{\cal L}_{\rm D, p}(\rho) = {\cal L}_{\rm D}(\rho) - i \frac{\Omega}{2} [S_x , \rho ]\,.
\label{Dicke.pumped}
\end{equation}
Competition between the collective decay and the pumping leads to a non-equilibrium 
phase transition in the steady-state of the model at a critical pumping rate $\Omega_{\rm c} = N \Gamma / 2$ \cite{Drummond78oc}, manifested in a kink in the population inversion observable $\mean{S_z}$, see Fig. \ref{fig2}(a). Let us first give a brief description of the two limiting cases:
{\it (i) Coherent steady state regime}, 
$\Omega \ll N \Gamma / 2$. Since ${\cal L}_{\rm D, p}$ can be obtained from 
${\cal L}_{\rm D}$ by the substitution $S^- \to S^- + i \Omega/(2 \Gamma)$, one can easily show that 
$\rho_s = | \Psi_\cc \rangle \langle \Psi_\cc | + {\cal O}^2(\frac{\Omega}{\Gamma})$, where $|\Psi_\cc \rangle= e^{i \frac{\Omega}{\Gamma} S_x} |N/2, -N/2 \rangle$ is a spin coherent state.
{\it (ii) Mixed state phase}, $\Omega \gg N \Gamma /2$. Here we get an infinite temperature state. To show it, it is convenient to write ${\cal L}_{\rm D}$ in the interaction picture with respect to $\Omega S_x/2$. This accounts to replace $S^- \to S_x + (1/2)(\cos(t) S_y + \sin(t) S_z)$. Averaging over time, leads to 
\begin{equation}
{\cal L}_{\rm D, p} \approx 
\frac{\Gamma}{2} 
\left(S_x \rho S_x - S_x^2 \rho  + 
\frac{1}{2} \sum_{\alpha = y,z} \left(S_\alpha \rho S_\alpha - S_\alpha^2 \rho \right) \right) + {\rm H.c.}\,. 
\end{equation}
which has the infinite temperature state $\rho_{\rm s} = {\bf 1}$ as steady-state. 
For calculations in the intermediate regime we use the full solution in the $| J, M \rangle$ basis.
\begin{figure}[h]
\begin{center}
\includegraphics[width=0.45\textwidth]{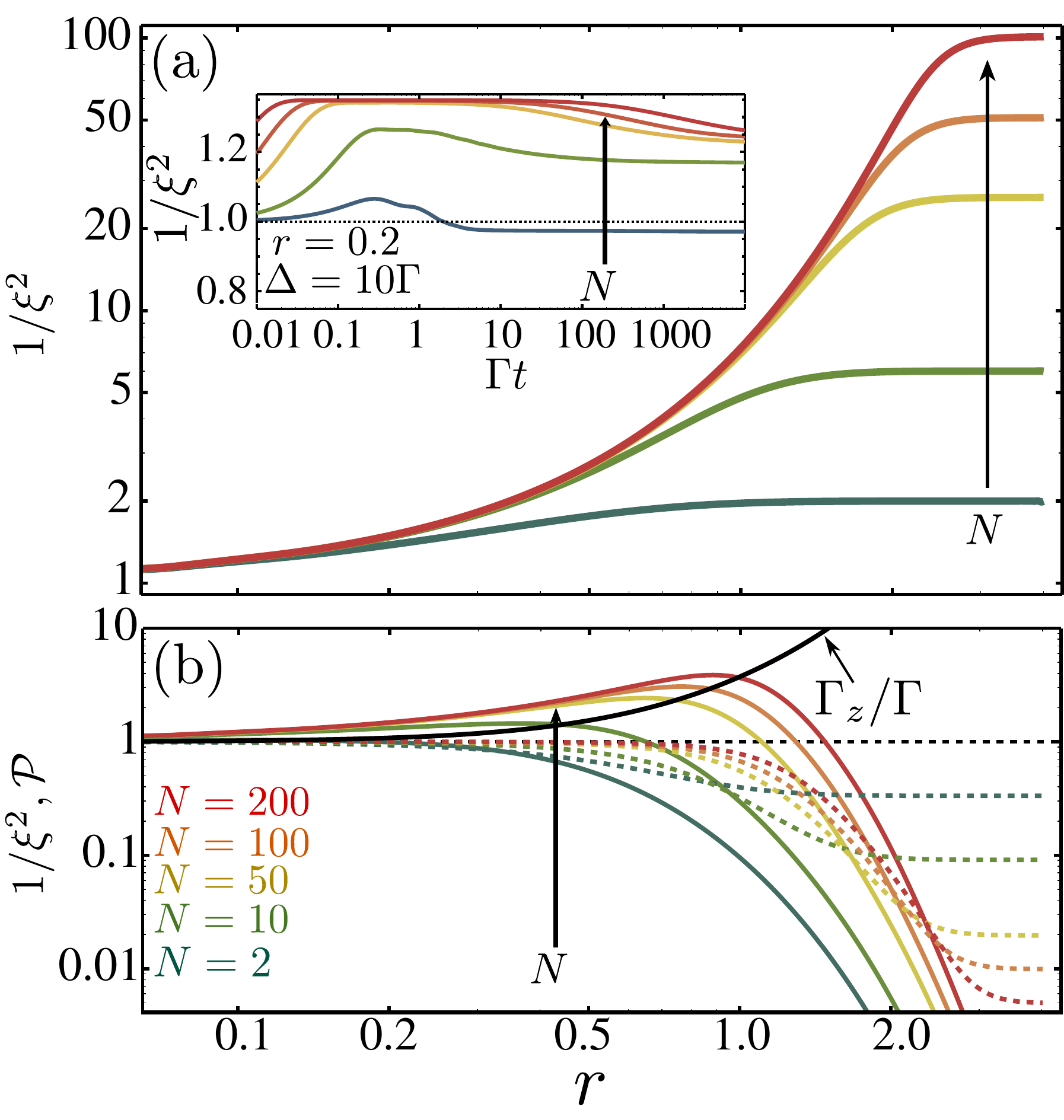}
\end{center}
\caption{(Color online) Same as Fig. \ref{fig2} for the 4LS scheme.  
(a) Entanglement witness ($1/\xi^2$,solid) as a function of the squeezing parameter, $r$, for increasing number of qubits ($N=2,10,50,100,200$) in the limit of {\it small photonic bandwidth} limit, where  $J_{m,n}^z\equiv 0$. In all the cases, the purity of the system is $\mathcal{P}\approx 1$. Inset: Dynamics of the entanglement witness  ($1/\xi^2$) for the ensembles of qubits with a fixed random dispersion of qubit energies, $\Delta=10\Gamma$, for the different number of qubits depicted in the main panel. (b) Entanglement witness  ($1/\xi^2$, solid) and purity ($\mathcal{P}$, dashed) as a function of the parameter $r$ for increasing number of qubits ($N=2,10,50,100,200$) in the limit of {\it large bandwidth} limit. The evolution of the collective dephasing mechanism, $\Gamma_z$, with the squeezing parameter is also plotted in solid black.} 
\label{fig3}
\end{figure}

To quantify entanglement we use spin-squeezing $\xi$ as a figure of merit,
\begin{equation}
\xi^2 = \frac{N (\Delta S_x)^2}{\langle S_y \rangle^2 + \langle S_z \rangle^2}\,.
\end{equation}
The latter is both an entanglement witness and it is also linked to applications in quantum metrology \cite{Wineland94pra, Sorensen01nat,Kitagawa93pra}.
Symmetric multiqubit states with $\xi < 1$ can be shown to be entangled 
\cite{Sorensen01nat}
with pairwise entanglement between any pair of qubits \cite{wang03a}. Note that the above mentioned (i) and (ii) phases lead to $1/\xi^2 = 1$ and $1/\xi^2 = 0$, respectively. Another theoretical tool to be used is the purity, defined by ${\cal P}(\rho) = \TTr{(\rho^2)}$, although we note that $\xi$ witnesses entanglement also for mixed states. 
Both magnitudes are plotted for increasing number of qubits $N$ in Fig. \ref{fig2}. 
We find a range of pumping fields ($\Omega \leq \Omega_{\rm c}$) that induce a pure entangled $\rho_{\rm s}$. This result leads to the controllable generation of entangled states in mesoscopic samples of artificial atoms. The timescale needed to achieved the stationary entangled states benefits from a collective enhancement scaling as $\Gamma N$, so that the higher the number of qubit, the more efficient the preparation of states is.

We upgrade now our 2LS into a 4-level system (4LS) configuration, see Fig. \ref{fig1} (c), and show that this is a more advantageous situation. Our scheme can  be realized in the solid-state context \cite{Elzerman11a, Weiss12prl} and describes a variety of possible configurations in which a set of low-level states are coupled to excited states by lasers with different polarizations.
Two ground states ($\ket{g}_n, \ket{e}_n$) are coupled to high energy states 
($\ket{g'}_n$, $\ket{e'}_n$). The qubit part of the free Hamiltonian becomes now
$H_{\rm qb} = 
\sum_n (
\omega_{g'} |g'\rangle_n \langle g'| + 
\omega_{e'} |e'\rangle_n \langle e'| + 
\omega_{g}  |g\rangle_n \langle g|)$. Two weak non-resonant fields with amplitudes $\Omega_{\LL,1(2)}$ and frequencies $\omega_{\LL,1(2)}$, 
induce transitions described by a Hamiltonian term
$H_{\LL} = 
\sum_{n} ( 
(\Omega_{\LL,1}/2)  |e'\rangle_n \langle e| e^{-i \omega_{\LL,1}t} + 
(\Omega_{\LL,2}/2)  |g'\rangle_n \langle g| e^{-i \omega_{\LL,2}t} + 
{\rm H.c.})$. We impose the condition $\omega_{\LL,1} - \omega_0 = \omega_{\LL,2} + \omega_0=\omega_a$, such that the two decay channels in red of Fig. \ref{fig1}(c) into modes $a_q$, correspond to photon emission with the same energy $\omega_a$.
After an adiabatic elimination of the excited states (see Sec. II  of Supp. Mat. for details) we get  an effective qubit-field interaction 
$H_{\II}(t) = H_{\II}^{\rm sq}(t) + H_{\II}^{z}(t)$, 
written in the interaction picture with respect to $H_0$. The first term reads
\begin{equation}
 \label{hamNgen}
H_{\II}^{\rm sq}(t)
=  \sum_{n} \kappa E(x_n,t)(D_n^\dagger e^{i\omega_a t} + D_n e^{-i\omega_a t})\,,
\end{equation}
where $\kappa^2 = 
(\Omega_{\LL,1}/2 \Delta_1)^2 - (\Omega_{\LL,2}/2 \Delta_2)^2$ 
($\Delta_{1(2)} = \omega_{e'(g')} - \omega_{\LL,1(2)}$) is a normalization constant.
$D_n = u\sigma^-_n + v\sigma^+_n$ is a jump operator resulting from the cross radiative decay, with $u = \kappa^{-1} \Omega_{\LL,1}/2\Delta_1$, 
$v = \kappa^{-1} \Omega_{\LL,2}/2\Delta_2$, fulfilling $u^2 - v^2 = 1$. The latter condition will be useful in the discussion below, and allows us characterization by a single parameter $r$, such that 
$u = \cosh(r)$, $v = \sinh(r)$. 
After eliminating the e.m. field degrees of freedom we arrive to the Liouvillian 
\begin{equation}
\label{mequation2}
{\cal L}_{\rm sq}(\rho) = 
\sum_{n,m} J^{\rm sq}_{n,m} \left( D^-_n \rho D^+_m - \rho D^+_m D^-_n \right)
+ \rm {H.c.}\,,
\end{equation}
with $J^{\rm sq}_{n,m} = \Gamma_{\rm sq} e^{i q(\omega_a) |x_n - x_m|}$ and
$\Gamma_{\rm sq} = \kappa^2 \gamma(\omega_a)$. 
The second term in the effective qubit-field interaction describes longitudinal decay processes,
\begin{equation}
 \label{hamNgensigmaz}
H_{\rm I}^z(t)
=\kappa \sum_n E(x_n,t)(u\sigma^{z}_n e^{i\omega_1 t}+v\sigma^{z}_n e^{-i\omega_2 t}+ \rm H.c.)\,,
\end{equation}
and leads to
\begin{equation}
{\cal L}_{z}(\rho) = 
\sum_{n,m} J^{z}_{n,m} \left( \sigma^z_n \rho \sigma^z_m - \rho \sigma^z_m \sigma^z_n \right)
+ \rm {H.c.}\, ,
\label{mequation3}
\end{equation}
with
$J^z_{n,m} = 
\kappa^2 
(\gamma(\omega_1) u^2 e^{i q(\omega_1) |x_n - x_m|} + \gamma(\omega_2) v^2 e^{i q(\omega_2) |x_n - x_m|})$ where $\omega_1=\omega_{\LL,1}$ and $\omega_2=\omega_{\LL,2}-\omega_0$. The term ${\cal L}_z$ induces a dephasing mechanism that competes with the spontaneous coherence build up induced by ${\cal L}_{\rm sq}$. The relative importance of those contributions depends on the photon density of states at frequencies $\omega_{\rm a}$ and 
$\omega_{1,2}$. We consider two limiting cases:
{\it (i) Small photonic bandwidth.-} 
This is the most favorable configuration. 
We assume that the density of states in the waveguide is peaked around 
$\omega_{\rm a}$, 
with a bandwidth $\Delta \omega \ll |\omega_1-\omega_{\rm a}|, |\omega_2-\omega_{\rm a}|$ 
such that 
$\gamma(\omega_{1,2})\approx 0$ and therefore $J^z_{n,m}\approx 0$. For example, this can be the case
of one dimensional waveguides consisting of coupled cavities forming a one-dimensional photonic crystal. 
Defining $q(\omega_a) =2 \pi /\lambda_a$ and choosing $x_n = n \lambda_a$ we arrive to a spin-squeezed version of the Dicke superradiant model, 
\begin{eqnarray}
{\cal L}_{\rm sq, D}(\rho) &=& 
\frac{\Gamma_{\rm sq}}{2} \left( D^- \rho D^+ - D^+ D^- \rho  + {\rm H.c.} \right)\,,
\label{collectivesqueezing}
\end{eqnarray}
where we have introduced collective spin-squeezed operators $D^{+/-} = \sum_n D_n^{-/+}$. In Fig. \ref{fig3} (a) we present a calculation of the spin-squeezing in the steady-state as function of the squeezing parameter, $r$. Remarkably, we observe an enhancement of the maximum value of entanglement of several orders of magnitude compared to the case of an ensemble of 2LS's.
{\it (ii) Large photonic bandwidth.-}
In the opposite limit we consider a broadband waveguide \cite{martincano10a,Gonzalez11prl} 
($|\omega_{{\rm L},1} - \omega_{{\rm L}, 2}|\ll \Delta\omega$) such that the density of states at the frequencies considered here is comparable $\gamma(\omega_1)\approx \gamma(\omega_2)$. 
In experiments with optical transitions, for example with quantum dots in optical or plasmonic waveguides, condition $\omega_1, \omega_2, \omega_a \gg \omega_0$ is found, since transition energies are in the $eV$ and $meV$ for high energy and low energy transitions, respectively \cite{Elzerman11a,Weiss12prl}. Thus, we can safely assume $q(\omega_a) \approx q(\omega_1) \approx q(\omega_2) = 2 \pi /\lambda_a$, and consider that quantum dots can be placed at the same relative optical path with respect to all frequencies. To give a more quantitative argument for this approximation we define the group velocity of the modes of the waveguide
$v_g(\omega)=|\partial_q \omega_q|$, and consider the limit
$|\omega_1 - \omega_{\rm a}| v_g(\omega_{\rm a}),
|\omega_2 - \omega_{\rm a}| v_g(\omega_{\rm b}) \ll q(\omega_{\rm a})$, which corresponds to small wavevector differences. In the case of constant $v_g$ and optical transitions, this condition leads to differences of $10^{-3}$ in 
$q(\omega_{\rm a})$, 
$q(\omega_{1,2})$. We neglect for the moment those differences, which may lead to inhomogeneous broadening effects that are discussed later in this work. Thus, he condition $x_n = n \lambda_a$ leads to a collective dephasing term of the form
\begin{eqnarray}
{\cal L}_{\rm z, D}(\rho) &=& 
\frac{\Gamma_z}{2} \left(S^z \rho S^z -  S^z \rho  + {\rm H.c.} \right)\,,
\label{collectivesqueezingdephasing}
\end{eqnarray}
where we have introduced the rate $\Gamma_z = J^z_{n,n}$. In the large photonic bandwidth limit we get thus two competing terms 
${\cal L} = {\cal L}_{\rm sq, D} + {\cal L}_{\rm z, D}$.
Collective dephasing increases with the squeezing parameter $r$, as depicted in solid black in Fig. \ref{fig3} (b). The competition between dephasing and squeezing mechanisms determines an optimal $r$ to generate maximal entanglement. 
The latter can be higher than the one generated by ${\cal L}_{\rm D, p}$ in the 2LS scheme considered above. The large bandwidth limit is a worst-case scenario as typically the waveguides mode are peaked around a certain value chosen by fabrication. Thus, in the realistic case the entanglement generation will be in a situation between the two limits. The purity of the system is also affected by the dephasing term, however one can still find a region that combine high purity and high values of entanglement as shown in Fig. \ref{fig3} (b).

Finally, we discuss the feasibility of our ideas focusing on the following points: 
{\it (i) One-dimensional waveguides.-}
We require long propagation length and efficient coupling to the qubits. Coupling to guided modes of $85-89 \%$ has been reported for photonic \cite{lundhansen08a,laucht12a} and plasmonic \cite{Akimov07nat,huck11a} waveguides. Theoretical predictions of even higher efficiencies has been pointed out \cite{lecamp07a,martincano10a}, although at the expense of reducing the field propagation length. Besides the precise location of the qubits is also required, which is possible for solid-state emitters using, i.e, lithographic methods which nowadays have precision larger than $50 nm$ \cite{mohan10a}. 
{\it (i) Lambda-transitions in solid-state qubits.-}
We assume a degree of addressability of electronic levels similar to the one achieved in atomic physics, specially the 4LS scheme. Applications in quantum information processing \cite{Imamoglu99prl,Gammon07prl} typically require controlling optical transitions for spin-pumping and initialization. Recent experimental results \cite{Elzerman11a, Weiss12prl} show level schemes in quantum dots similar to those required in our work.
{\it (iii) Markovian approximation.-}
We require $\Gamma N \ll \omega_0$ in the 2LS scheme and 
$\Gamma N \ll \omega_{\rm L 1,2}, \omega_{\rm a}$ in the 4LS case, such that the cooperative decay rate is much smaller than the transition frequencies, the latter determine the photonic bath memory time \cite{breuerbook}. This condition is well satisfied in the case of 
optical transitions of quantum dots.
{\it (iv) Independent decay channels on each transition frequency.-} This is required for the 4LS's scheme in the large bandwith limit, to get Eqs. (\ref{mequation2}, \ref{mequation3}), and it is justified as long as
$\Gamma N \ll |\omega_{\rm L,1} - \omega_{\rm a}|, 
|\omega_{\rm L,2} - \omega_{\rm a}|$. Since differences in transition energies are of the order of $meV$, this condition imposes a restriction on the achievable rates for entanglement generation in our scheme.
{\it (v) Homogeneous couplings.-} So far we have neglected inhomogeneities in the couplings and qubit energies, which take the steady state out of the $\ket{J,M}$ basis. 
This may be a severe restriction in quantum optical solid-state devices. Although inhomogeneous broadening in solid-state setups is still of the order of $meV$ for quantum dots \cite{mohan10a} and  $\mu eV$ for NV centers \cite{kubo12a}, the feasibility of our proposal will benefit from current experimental efforts in the field.

We have carried out calculations to check the effect of experimental imperfections with a focus on an inhomogeneous random distribution of qubit energies, $\Delta \omega_j$, described by  
a term $H_{\rm inh} = \sum_j \Delta \omega_j \sigma^z_j$ (with $\Delta \omega_j\in [-\Delta, \Delta]$). 
Exact calculations are very demanding, however for a limited number of qubits $N = 2,3,4$, we are able to show that $H_{\rm inh}$ induces a dephasing time $t_{\rm d}$, such that for $t > t_{\rm d}$,
spin-squeezing is totally degraded in the 2LS scheme, or strongly decreases below its maximum value in the 4LS case (see \cite{SM} for details). 
In the 4LS scheme, one can use a bosonic approximation in the master equation ($\sigma_n \approx b_n$, with $b_n$ a bosonic annihilation operator) and render the problem solvable in a low occupation limit 
$\langle \sigma^+_n \sigma_n \rangle \approx 0$. 
This method has allowed us to study the scaling of the spin-squeezing for large $N$. 
Our main result is that, under the effect of $H_{\rm inh}$, the system reaches the steady-state spin-squeezing values, and after a time $t_d$, entanglement degrades down to a residual value. The robustness of the 4LS scheme increases for large $N$, since $t_{\rm d}$ grows with $N$. In Fig. \ref{fig3}(a) (Inset) we show results for small values of $r$, which are particularly well described by the bosonization method. We confirm the same scaling with larger values of $r$ with higher degrees of $1/\xi^2$ (see \cite{SM} for details). Our conclusion is that the 4LS is advantageous with respect to the 2LS, since it allows us to generate higher spin-squeezing values with $t_{\rm d}$ increasing with $N$.

In conclusion, we have proved that one-dimensional plasmonic \cite{Akimov07nat,huck11a} or photonic \cite{lonvcar00a,vlasov05a,schwoob05a,lundhansen08a,laucht12a} waveguides can be used to correlate a large number of qubits by collective radiative decay. Our scheme is feasible in solid-state devices currently under investigation for quantum information processing. Those ideas can be translated to circuit QED by controlling the qubit-field coupling (see \cite{Porras12prl}).
During completion of this work we became aware of a theoretical preprint on atomic ensembles in single-mode optical cavities \cite{dallatorre12a} related to our 4LS scheme. In our work we assume the continuum of modes in a one-dimensional waveguide, and thus we do not require energetically resolving a single cavity mode, which would hinder scaling up our scheme in solid-state setups with large number of qubits

{\it Acknowledgments.-} We acknowledge QUITEMAD S2009-ESP-1594, MICINN-MAT2011-22997,CAM- S-2009/ESP-1503, FIS2009-10061, CAM-UCM/910758, RyC Contract Y200200074, and FPU grants AP2008-00101. We thank to J. Miguel-Sanchez for useful discussion of the experimental conditions.

\bibliography{references_diego}
\end{document}